\documentclass{easychair}
\usepackage[english]{babel}

\usepackage{amsmath}
\usepackage{graphicx}
\usepackage{comment}
\usepackage{natbib}
\usepackage{caption}
\usepackage{subcaption}
\usepackage{pdflscape}
\usepackage{tabularx,booktabs,array, pifont, ragged2e, enumitem}

\usepackage{float}
\usepackage{algorithm}
\usepackage{algpseudocode}   
\usepackage{stfloats,placeins}

\usepackage{xspace}
\newcommand{\ie}{i.e.\@\xspace} 
\newcommand{\eg}{e.g.\@\xspace} 

\usepackage{url}
\usepackage{stfloats}   

\newcommand{\cmark}{\ding{51}} 
\newcommand{\xmark}{\ding{55}} 

\AtBeginDocument{%
  }
\usepackage[acronym, nomain]{glossaries}
\glsdisablehyper

\newacronym{apk}{APK}{Android Package}
\newcommand{\APK}{\gls{apk}\xspace}

\newacronym{os}{OS}{Operating System}

\newacronym{vm}{VM}{Virtual Machine}

\newacronym{dvm}{DVM}{Dalvik Virtual Machine}

\newacronym{jvm}{JVM}{Java Virtual Machine}

\newacronym{art}{ART}{Android Runtime}

\newacronym{aot}{AOT}{Ahead-of-Time}

\newacronym{jit}{JIT}{Just-in-Time}

\newacronym{adb}{ADB}{Android Debug Bridge}

\newacronym{abi}{ABI}{Application Binary Interface}

\newacronym{df}{DF}{Digital Forensics}
\newcommand{\DF}{\gls{df}\xspace}

\newacronym{ai}{AI}{Artificial Intelligence}
\newcommand{\AI}{\gls{ai}\xspace}

\newacronym{ml}{ML}{Machine Learning}

\newacronym{dl}{DL}{Deep Learning}

\newacronym{cnn}{CNN}{Convolutional Neural Network}

\newacronym{llm}{LLM}{Large Language Model}
\newcommand{\LLM}{\gls{llm}\xspace}

\newacronym{nlp}{NLP}{Natural Language Processing}
\newcommand{\NLP}{\gls{nlp}\xspace}

\newacronym{xai}{xAI}{eXplainable AI}

\newacronym{cti}{CTI}{Cyber Threat Intelligence}

\newcolumntype{Y}{>{\RaggedRight\arraybackslash}X}

\title{An Explainable Memory Forensics \\ Approach for Malware Analysis}

\author{
    Silvia Lucia Sanna\inst{1}
\and
    Davide Maiorca\inst{1}
\and
    Giorgio Giacinto\inst{1}\inst{2}
}

\institute{
  University Of Cagliari,
  Italy\\
  \email{name.surname@unica.it}
  \and
  Consorzio Interuniversitario Nazionale per l’Informatica (CINI), 
  Italy
}

\institute{
  University Of Cagliari,
  Italy\\
  \email{name.surname@unica.it}
  \and
  Consorzio Interuniversitario Nazionale per l’Informatica (CINI), 
  Italy
}

\titlerunning{Explainable Memory Forensics}

\authorrunning{Sanna et.al.}

\begin{document}
\maketitle

\begin{abstract}
Memory forensics is an effective methodology for analyzing living-off-the-land malware, including threats that employ evasion, obfuscation, anti-analysis, and steganographic techniques. By capturing volatile system state, memory analysis enables the recovery of transient artifacts such as decrypted payloads, executed commands, credentials, and cryptographic keys that are often inaccessible through static or traditional dynamic analysis. While several automated models have been proposed for malware detection from memory, their outputs typically lack interpretability, and memory analysis still relies heavily on expert-driven inspection of complex tool outputs, such as those produced by Volatility. In this paper, we propose an explainable, AI-assisted memory forensics approach that leverages general-purpose large language models (LLMs) to interpret memory analysis outputs in a human-readable form and to automatically extract meaningful Indicators of Compromise (IoCs), in some circumstances detecting more IoCs than current state-of-the-art tools. We apply the proposed methodology to both Windows and Android malware, comparing full RAM acquisition with target-process memory dumping and highlighting their complementary forensic value. Furthermore, we demonstrate how LLMs can support both expert and non-expert analysts by explaining analysis results, correlating artifacts, and justifying malware classifications. Finally, we show that a human-in-the-loop workflow, assisted by LLMs during kernel-assisted setup and analysis, improves reproducibility and reduces operational complexity, thereby reinforcing the practical applicability of AI-driven memory forensics for modern malware investigations.

\end{abstract}

\section{Introduction}
Malware analysis is currently facing different challenges because of the development of stealthy malware \eg evasive, obfuscated, stegomalware~\cite{Ruggia24_AndroidEvasive, Soi_IHMMSEC25}, both for Windows desktop computers and Android mobile devices, the most used Operating Systems worldwide. Traditional techniques involving static (\ie without execution) and dynamic (\ie runtime monitoring) analysis are sometimes inefficient for such specific malware classes. Firstly, obfuscated malware cannot be detected using static analysis techniques, especially when using pattern matches (\ie regex) and the API names are not in clear. Even if static and dynamic methodologies have been combined during the years with \AI algorithms \cite{Apruzzese2022}, they lost accuracy with the development of adversarial samples, \ie malware evading \AI-based detectors~\cite{Demontis19_IEEE}. For this reason, memory analysis is emerging in recent studies~\cite{Ali-Gombe19_RAID, Aligombe_DFRWS23_crgbmem, Khalid_volmemdroid_ESA24, Bellizzi_Access22, Bellizzi_JCP23_vedrando}. They all demonstrated that memory forensics enables the extraction of specific artifacts that are not retrievable with other techniques. Current works provide different acquisition techniques and different analysis strategies. Acquisition methodologies are divided into target-application analysis and full memory extraction. The acquisition methodology depends on the target OS. For Android, full memory acquisition requires the kernel recompilation in order to load the {\tt LiME} (Linux Memory Extractor)~\cite{504ensicsLabsLiME} kernel module, the only available tool for complete memory dumping. This approach is helpful to study the effects of the malware on the whole system by analyzing the interaction with the OS and other processes, file access and creation, environmental variable, network traffic, bash commands. On the contrary the target-app dump in Android uses Fridump, based on instrumentation tools such as {\tt Frida}~\cite{yang2019fridump, Frida}, only extracts the memory allocated to the target process under analysis. For Windows, the complete memory can be extracted with specific tools like {\tt FTK-Imager}, or {\tt VirtualBox Management} when using a Virtual Machine. On the other hand, target-software application can be extracted with tools like {\tt procdump}.

The analysis methodologies are based on specific content extraction such as allocated objects~\cite{Ali-Gombe19_RAID}, binary-to-image encoding~\cite{Aligombe_DFRWS23_crgbmem}, specific IoC matching through strings and regex matching~\cite{Khalid_volmemdroid_ESA24, bellizzi2022jitmf, Bellizzi_SIP21}. However, none of the current studies focused on the interpretability of the complete memory extraction and subsequently the full analysis of its content for malware analysis. They merely rely on specific known pattern matching~\cite{Khalid_volmemdroid_ESA24}. Additionally, to the best of our knowledge, there is no study based on the differences between the complete memory dump and the target-app dump. Regarding the complete memory analysis, the only available tool is {\tt Volatility} which requires a deep knowledge on its usage and data interpretability. We claim that general-purpose LLMs can be applied in these scenarios even for teaching purposes by giving a human-readable interpretation of the tool output \cite{Mengqi24_ACM, zhao2025surveylargelanguagemodels}. This concept has already been explored in traditional teaching but to the best of our knowledge there is still no application in \DF.

In this study, we propose a comparative analysis of the full RAM extraction and the target-process dump. In our case, the analysis is guided by general-purpose LLMs and \NLP algorithms. In fact, by analyzing both the {\tt Volatility}~\cite{VolatilityFramework} outputs and the text encoding (\ie reading the strings of the dump) from the memory extracted using both approaches, we can give to the human analyst a comprehensive report on the interpretation of \emph{(i)} the {\tt Volatility} outputs, \emph{(ii)} the details of the found artifacts and \emph{(iii)} an explanation according to the given classification of the algorithm. If the memory is captured over time, our approach can also evaluate the differences in the dumps evolving during runtime app execution. We opt for general-purpose LLMs to adapt their vast ground truth for detection (\ie, an \LLM like ChatGPT would be more efficient than a self-hosted \LLM trained on a few under-representative data). In fact the dataset is a huge problem in \DF research and investigations, that because of the lack of public databases, cases, files and evidences, research goes slightly slow~\cite{Breitinger2024DFRWS10yr, gobel2023data}. Additionally, real cases cannot be used if not properly managed according to privacy preservation~\cite{Mombelli2024FAIRness, gobel2023data}. In this way, synthetic datasets are a plausible solution. Our analysis is supported by the release of a unified methodology for full RAM extraction. We ask the following research questions:
\begin{itemize}
    \item \textbf{RQ1}: Can AI be applied as an assistant in Digital Forensics analysis?
    \item \textbf{RQ2}: What is the difference between complete memory and target-process analysis?
    \item  \textbf{RQ3}: Can general-purpose LLMs help the analyst in malware detection?
\end{itemize}

Our main contributions are threefold. First, we introduce an explainable, AI-assisted memory forensics pipeline that leverages general-purpose LLMs to interpret volatile memory artifacts and analysis outputs in a human-readable and forensically meaningful manner. Second, we provide a systematic comparative analysis between full RAM acquisition and target-process memory dumping, highlighting their differences in malware visibility, forensic coverage, and the evolution of malicious behavior over time during execution. Third, we explore and demonstrate the effectiveness of general-purpose LLMs as analysis assistants for malware memory forensics, showing how they can support both expert and non-expert analysts in understanding complex memory artifacts. In addition, we address the practical challenges of volatile memory acquisition by adopting a human-in-the-loop methodology during the kernel-assisted setup phase. To streamline this complex and error-prone process, we used ChatGPT as an interactive assistant, requesting step-by-step guidance tailored to the target kernel and iteratively providing compilation and runtime error logs to resolve toolchain mismatches, missing headers, module loading failures, and symbol extraction issues. This approach significantly reduced setup time while preserving methodological transparency, as all corrective actions were driven by explicit error traces. To further support reproducibility and independent validation, we release the compiled kernel artifacts and associated build metadata used in our experiments. Finally, we emphasize the critical role of memory forensics in modern malware detection by comparing memory-based analysis results with traditional approaches, such as VirusTotal~\cite{virustotal}, Drebin~\cite{drebin}, and Entroplyzer~\cite{Keyes21_RDAAPS}, thereby illustrating the added value of volatile-memory analysis against stealthy and evasive malware.

The remainder of this paper is structured as follows. Section~\ref{sec:backgroundsota} highlights the main concepts for memory forensics and the current techniques used in malware analysis, while Section~\ref{sec:approach} introduces the used Methodology. The memory analysis results of Windows and Android using LLMs and \NLP are presented respectively in Section~\ref{sec:resultswin} and Section~\ref{sec:resultsandro}. Section~\ref{sec:conclusions} closes the paper answering to the previously posed Research Questions.
\label{sec:intro}

\section{Background and Related Works}
Memory forensics is a fundamental component of modern malware analysis, enabling investigators to observe malicious behavior at runtime and recover evidence that is deliberately concealed from persistent storage. Unlike static disk-based analysis, memory forensics captures the live execution state of a system, exposing transient artifacts such as decrypted malware payloads, unpacked shellcode, injected dynamic libraries, reflective loaders, credential material, encryption keys, command-and-control configurations, and active network sessions. These properties make memory analysis particularly effective against fileless malware, packers, droppers, and multi-stage attacks, which increasingly dominate both Windows and Android threat landscapes \cite{Bhusal_ACM25, gaber_malware_2024}.

\textbf{Memory Forensics for Windows Malware Analysis}
Windows memory forensics represents the most mature and operationally deployed domain of volatile analysis. Early foundational work demonstrated that malware frequently resides exclusively in memory, motivating the systematic analysis of RAM to detect rootkits, injected code, and kernel-level manipulation \cite{Carvey2007,Schuster2006}. Modern Windows malware routinely employs runtime-only techniques such as process hollowing, reflective DLL injection, APC injection, direct system call invocation, and kernel rootkits, all of which leave minimal or no traces on disk \cite{Okolica2011,Gianni2016}. Memory acquisition on Windows is typically performed using kernel drivers or trusted execution environments, enabling full physical memory capture with high fidelity. Tools such as \texttt{WinPmem}, \texttt{DumpIt}, and \texttt{FTK Imager} are widely used in live-response scenarios, while hypervisor-based acquisition enables repeatable malware analysis in sandboxed environments \cite{Sylve2012,Case2013}. Compared to mobile platforms, Windows benefits from a more stable kernel architecture and symbol infrastructure, facilitating reliable reconstruction of kernel objects and process internals. Post-acquisition analysis is commonly conducted using frameworks such as \texttt{Volatility} and \texttt{Rekall}, which reconstruct malware-relevant artifacts including hidden processes, injected memory regions, anomalous virtual address descriptors (VADs), malicious kernel modules, registry hives, credential material, and in-memory PE files \cite{VolatilityFramework,rekall2017}. These techniques are central to the detection of advanced persistent threats (APTs), ransomware, and fileless malware campaigns that evade traditional signature-based defenses \cite{Okolica2011,Gianni2016}.

\textbf{Memory Forensics for Android Malware Analysis} Memory forensics has also become increasingly important for Android malware analysis, particularly in response to widespread use of dynamic loading, runtime unpacking, and native code exploitation. Android malware often decrypts payloads only at execution time, dynamically loads DEX files or native libraries, and manipulates ART or JNI components, behaviors that are frequently invisible to static APK analysis \cite{Ali-Gombe19_RAID, Aligombe_DFRWS23_crgbmem}. Android memory forensics inherits many principles from Linux but introduces additional complexity due to kernel fragmentation, device heterogeneity, and security mechanisms such as SELinux, Verified Boot, and application sandboxing \cite{Afonso2016}. As a result, full physical memory acquisition is less commonly achievable than on Windows systems, and many approaches rely on partial acquisition strategies, including process dumps, heap analysis, and runtime instrumentation \cite{Zhang18_CPS, bellizzi2022jitmf}. Despite these constraints, memory-based analysis enables the recovery of crucial malware artifacts, including decrypted DEX code, native shellcode, runtime-generated strings, cryptographic keys, and command-and-control endpoints \cite{Ali-Gombe19_RAID,Franzen22_ACM, Bellizzi_SIP21}. These capabilities are essential for understanding modern Android malware that actively evades static and network-based detection.

textbf{AI for Memory Forensics} The rapid growth in data volume and attack sophistication has driven the adoption of artificial intelligence (AI) and machine learning (ML) across digital forensics \cite{Garcia2018,Apruzzese2022}. ML techniques have been successfully applied to malware classification, anomaly detection, forensic triage, and artifact prioritization across disk, network, and memory data sources. Memory forensics, in particular, has emerged as a promising domain for AI-driven analysis due to the highly entropic and unstructured nature of RAM dumps. Statistical and ML-based approaches classify memory regions using features such as entropy, byte-frequency distributions, $n$-grams, and structural regularities, enabling detection of injected or packed malware components even when semantic parsing fails \cite{Keyes21_RDAAPS,Song2018}. Recent work explores representation learning to improve scalability and generalization. Image-based encodings transform memory or binary data into grayscale or RGB matrices suitable for convolutional neural networks, an approach initially proposed for Windows malware binaries \cite{Nataraj_VizSec11} and later extended to memory dumps and Android artifacts \cite{Aligombe_DFRWS23_crgbmem,Daoudi21_Springer}. These approaches enable featureless learning and visual explainability but may introduce artificial spatial correlations due to the reshaping of linear memory into two-dimensional layouts \cite{Malhotra2024_Virology}. Complementary approaches include sequence-based deep learning, entropy-flow modeling, and audio-based encodings of binaries \cite{Mercaldo2021,Kural2023}. However, most existing AI-based memory malware detection systems still operate on partial memory views or predefined regions of interest, limiting their ability to capture multi-stage or cross-process attacks. Explainability is increasingly recognized as a core requirement for forensic AI systems. In investigative and legal contexts, opaque predictions are insufficient; analysts must understand why a memory region is classified as malicious and how it relates to observable runtime behavior \cite{Ribeiro2016,Guidotti2019}. Explainable AI (xAI) techniques address this need by highlighting influential memory regions, byte patterns, or learned features that drive classification decisions. By integrating xAI into memory-based malware analysis, AI models transition from black-box classifiers to decision-support systems that expose actionable Indicators of Compromise (IoCs), such as suspicious memory segments, injected code regions, or anomalous runtime structures \cite{Keyes21_RDAAPS}. This capability is particularly valuable in volatile memory forensics, where evidence is ephemeral and must be interpreted rapidly and defensibly.
\label{sec:backgroundsota}

\section{Approach}
\begin{figure}
    \centering
    \includegraphics[width=\linewidth]{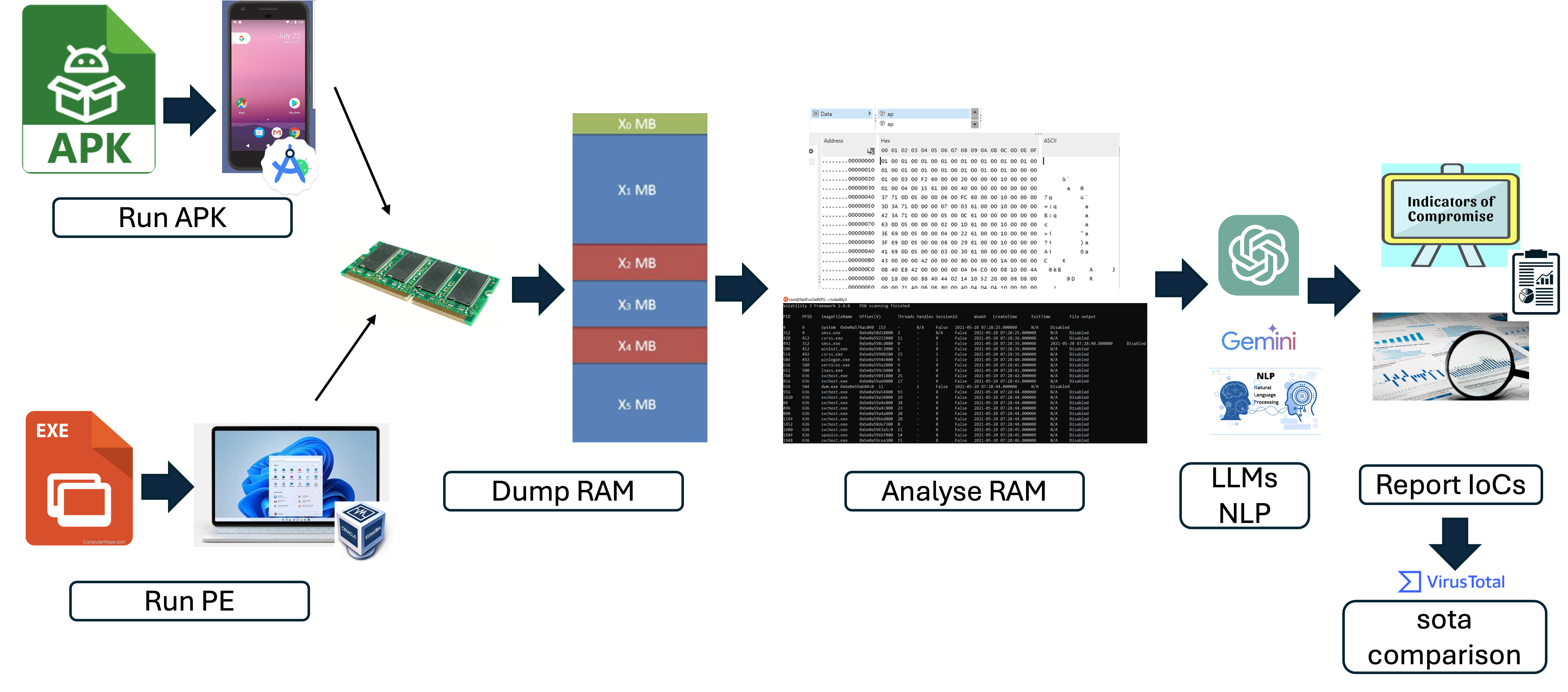}
    \caption{Execution pipeline. Run the target program (APK or PE) in the correct sandbox, extract the RAM related to the target process and the full RAM, analyse the dump and extract IoCs with LLMs and NLP algorithms, producing a human-readable report explaining the content of the dump, listing the found IoCs and explaining the classified executed program. The extracted results are compared with popular state-of-the-art tools}
    \label{fig:pipeline}
\end{figure}

In the following we show the detailed methodology used for this work. The pipeline reported in~\autoref{fig:pipeline} is designed to run a target program in the correct sandbox, extract the memory during execution, save the memory dump. If requested it is possible to repeat the execution over time, with a preferred time interval in order to have a sequential memory dump to study the behavior of the target malware sample during execution, sometimes with automatic or manual interaction to study the events. Subsequently, the dump is analysed with {\tt Volatility} in case of full RAM extraction, otherwhise with {\tt strings} to extract UTF-8 readable characters as no public tool is available for target-process memory analysis. At this point, we apply LLMs and NLP algorithms to extract specific IoCs from the dump, classify the target application with an explanation on the result. Additionally, we use the general-purpose LLMs to interpret and explain to the analyst the {\tt Volatility} output, a fundamental step for non-expert users. At the end we compare the extracted IoCs and classification report with common state-of-the-art tools to stress the power of AI-based memory forensics for malware analysis.

\subsection{Dataset Creation}
\label{sec:dataset}
We downloaded real malware samples for Android and Windows (the most used OS respectively for Desktop and Mobile) and executed in protected sandbox to not damage real environments as they are real malware. Despite some malware detect the presence of a debug environment, we excluded such samples otherwise the malware does not start. We considered dropper and fake APKs for Android because are one of the most dangerous categories to be detected. While for Windows we considered the ransomware as it is the main threat. We expect to extract important IoCs from the memory dump and improve the analysis of current state-of-the-art tools.

\subsection{Memory Extraction}
\label{sec:memoryextraction}
Due to the different architectures, the memory extraction is strictly dependant on the target OS. 

\textbf{Windows Memory Extraction} For Windows experiments, full RAM acquisition was performed using a hypervisor-assisted approach that does not require kernel instrumentation or elevated privileges within the guest operating system. All malware samples were executed inside a controlled VirtualBox virtual machine, and while the malware was running, the complete physical memory of the guest was captured directly from the host using the native \texttt{VBoxManage} command. This method exports a snapshot of the virtual machine’s RAM without interacting with the guest kernel or altering its runtime state, thereby ensuring acquisition transparency and minimizing observer effects. Because memory extraction is performed externally at the hypervisor level, this approach bypasses in-guest security mechanisms such as driver signing and kernel patch protection, while still providing a forensically complete memory image suitable for subsequent analysis with standard memory forensics frameworks.

\textbf{Android Memory Extraction} We adopted a kernel-assisted methodology for full RAM acquisition on Android, as complete volatile memory extraction necessarily requires execution in kernel context. Unlike desktop operating systems, Android enforces strict isolation between user space and kernel space through SELinux, application sandboxing, and Verified Boot, which prevents unprivileged processes from accessing physical memory or kernel data structures. As a consequence, super-user (root) privileges are mandatory to load kernel instrumentation and read physical page frames. Furthermore, common consumer and emulator kernels either disable loadable modules or enforce strict module signature verification, making third-party forensic modules incompatible with stock builds. For this reason, kernel recompilation is required to enable loadable module support, ensure ABI compatibility with the running kernel, and allow the insertion of the LiME acquisition module without altering runtime semantics. Our procedure follows the official Volatility Android guideline\footnote{\url{https://github.com/volatilityfoundation/volatility/wiki/Android}}, which prescribes target characterization, kernel source selection, configuration alignment, LiME out-of-tree compilation, and symbol/profile generation for memory analysis. To streamline this complex and error-prone process, we used ChatGPT as an interactive assistant, requesting step-by-step guidance tailored to the target kernel and iteratively providing compilation and runtime error logs to resolve toolchain mismatches, missing headers, module loading failures, and symbol extraction issues. This human-in-the-loop approach reduced setup time while preserving methodological transparency, as all corrective steps were driven by explicit error traces. Finally, to ensure reproducibility and facilitate independent validation, we release the compiled kernel artifacts and associated build metadata used in our experiments.

\subsection{Memory Analysis}
\label{sec:memoryanalysis}
In the following we present the memory analysis methodology details for full RAM extraction and target-process dumps.

\textbf{Full Memory Analysis} Memory analysis was performed using {\tt Volatility}, which requires kernel-specific metadata to correctly interpret raw memory images. While {\tt Volatility} automatically resolves symbols and analysis contexts for Windows memory dumps, Android memory analysis mandates the explicit creation of a kernel-matching profile due to the absence of standardized debugging symbols and the extensive fragmentation of Android kernels across versions and vendors. In this work, Volatility profiles were generated directly from the same custom kernel build used during memory acquisition, ensuring exact ABI consistency between the running system and the analysis framework. Specifically, kernel recompilation produced the unstripped {\tt vmlinux} binary and the corresponding {\tt System.map} file, which together encode symbol addresses and structure layouts. For Volatility~2, we extracted DWARF debugging information from {\tt vmlinux} to generate a {\tt module.dwarf} file and packaged it with {\tt System.map} into a profile archive, which was then loaded explicitly during analysis. For Volatility~3, the same kernel artifacts were processed using {\tt dwarf2json} to generate a JSON-based symbol file that encapsulates kernel types, offsets, and symbols in a structured format. This symbol file was supplied to Volatility at runtime to enable kernel and process reconstruction. The profile-generation step is mandatory on Android, as even minor kernel configuration changes alter structure offsets and invalidate generic assumptions. Although this procedure introduces additional complexity compared to Windows, it guarantees analytical correctness by ensuring that all memory objects are parsed using symbols derived from the exact kernel image responsible for the acquired RAM dump.

\textbf{Target Memory Analysis}
Process-level memory acquisition is performed using platform-specific tools. On Android, the memory of the target application was extracted using {\tt Fridump}, which attaches to the running process via Frida and dumps its mapped memory regions at runtime. On Windows, process and system memory acquisition was carried out using {\tt WinPmem}, a kernel-level acquisition tool that enables reliable extraction of volatile memory while the malware sample is executing. These targeted dumps complement full RAM acquisition by providing finer-grained visibility into application-specific runtime artifacts.

\subsection{AI-assisted Memory Analysis}
\label{sec:llm}
The main contribution of this paper is the application of general-purpose LLMs to analyse memory dumps while executing a malware. First of all, general-purpose LLMs are used to interpret the output of the analysis conducted in the complete memory dump with {\tt Volatility} as it produces outputs in table format, readable only by expert users and analysts. We apply LLMs to this output to help people understanding the output of this tool, hence for teaching improvements. Additionally, as general-purpose LLMs are trained on a large amount of data, we use them to extract automatically IoCs from the memory dump. This part is fundamental to understand if there is a malware running in the target device and because of memory forensics extract artifacts not detectable with traditional approaches as every data in RAM resides in plain text (\eg passwords, encryption keys). After extracting the IoCs, the LLM produces a report explaining the {\tt Volatility} output, the extracted IoCs and the classification of the program running on RAM with its explanation. We also compared the LLMs results with ad-hoc NLP algorithms and for target-application memory dump.
\label{sec:approach}

\section{Evaluation Results and Discussion }
In this set of experiments, we perform full RAM analysis following the methodology described in Section~\ref{sec:approach} and illustrated in Figure~\ref{fig:pipeline}. Memory dumps are examined using {\tt Volatility} version~2 (v2.6.1) and version~3 (v2.27.0), which extract forensic artifacts such as process trees, network connections, loaded libraries, memory mappings, and file system activity through their respective plugins. The analysis workflow is common to both Windows and Android: once memory is acquired, Volatility parses kernel and process structures and produces structured outputs that must be interpreted to assess malicious behavior.

To evaluate the difference between full RAM acquisition and target-process memory dumping, we designed experiments that repeatedly capture memory during execution. For each sample, memory was dumped five times within the first 60 seconds of execution (i.e., one dump every 10 seconds), allowing us to study the temporal evolution of malicious behavior. Full RAM analysis provides a system-wide view of attack effects, including file access, network activity, inter-process interaction, and injected code, while target-process dumping offers a more focused perspective on application-specific behavior. This experimental design is applied consistently across both operating systems, although the acquisition mechanisms differ by platform.

After acquisition, all memory dumps were parsed using a predefined set of Volatility plugins (reported in Table~\ref{tab:volplugins} for both Volatility~2 and Volatility~3), and their outputs were stored for subsequent analysis. Because Volatility produces large, table-based outputs that typically require expert interpretation, we investigated whether general-purpose LLMs, specifically ChatGPT-4o-mini and Gemini~2.0-flash-lite, could assist analysts in understanding these results. The prompts (reported in Table~\ref{tab:fullprompts}) instruct the LLMs to \emph{(i)} read Volatility plugin outputs, \emph{(ii)} translate them into human-readable explanations, \emph{(iii)} extract relevant Indicators of Compromise (IoCs), \emph{(iv)} correlate evidence across plugins to identify suspicious or malicious processes, and \emph{(v)} generate a final report explaining the classification decision.

In parallel, we implemented a custom rule-based NLP pipeline to analyze both Volatility outputs and raw strings extracted from memory dumps. This NLP component relies on deterministic regular-expression matching and entropy-based heuristics to detect forensic clues such as command execution patterns, network indicators, privilege abuse, code loading, and obfuscation artifacts. The resulting features are aggregated into interpretable vectors and classified as malicious, suspicious, benign, or unknown. Unlike embedding-based or statistical NLP approaches, this system is fully explainable and knowledge-driven, enabling direct attribution of decisions to specific textual evidence. The comparison between LLM-based interpretation and deterministic NLP analysis is summarized in Figure~\ref{fig:comparison}.
\label{sec:results}

\subsection{Windows}
The Windows experiments were conducted in a controlled virtualized environment using \texttt{VirtualBox}. 
A Windows~$11$ guest virtual machine was executed on the host system, and full memory acquisition was performed externally via \texttt{VBoxManage guestcontrol}. 
This approach allows reliable and repeatable memory snapshots without requiring kernel drivers or in-guest acquisition tools, thereby minimizing interference with the system under analysis.

\textbf{Memory Extraction}. For each experiment, the full RAM image was collected at runtime and subsequently analyzed using \texttt{Volatility~3}. 
Unlike Android experiments, no kernel recompilation or profile tuning was required, as {\tt Volatility~3} natively supports modern Windows memory structures. 
Each memory dump corresponds to a single execution snapshot, ensuring consistency between forensic artifacts and observed runtime behavior. No analysis could be conduct with {\tt Volatility~2} because when running the plugins to extract data from the memory dump a profile is required and no match is available for Windows~$11$. 

\textbf{Memory Analysis}.
The forensic analysis on Windows memory dumps focused on interpreting the outputs of a comprehensive set of \texttt{Volatility~3} plugins (shown in Table~\ref{tab:volplugins_windows}), including process enumeration, memory allocation structures, loaded modules, kernel callbacks, registry artifacts, and persistence-related mechanisms.
These plugins provide low-level visibility into system behavior that cannot be reconstructed from static binaries alone. The analysis was performed using three approaches: 
\emph{(i)} a rule-based heuristic NLP engine, \emph{(ii)} general-purpose Large Language Models (ChatGPT-4o-mini and Gemini~2.0-flash-lite), and \emph{(iii)} external reputation-based analysis via \texttt{VirusTotal}. Each approach was evaluated on the same set of Windows memory dumps to ensure comparability. A key advantage of LLM-based analysis lies in its ability to transform low-level forensic outputs into human-readable explanations. 
Both ChatGPT and Gemini were able to contextualize complex \texttt{Volatility} plugin outputs, such as suspicious memory regions, anomalous pool allocations, hidden processes, or injected code segments, and explain their forensic relevance in natural language. In contrast, the heuristic NLP approach produces interpretable but limited explanations based on predefined rules and regex-based IoC extraction. 
While effective at highlighting known patterns (\eg URLs, IP addresses, encoded blobs), it lacks the contextual reasoning required to relate multiple artifacts across different plugins. Overall, LLMs significantly improve forensic explainability by bridging the semantic gap between raw memory artifacts and analyst-level understanding.

\begin{figure}
    \centering
    \includegraphics[width=0.6\linewidth]{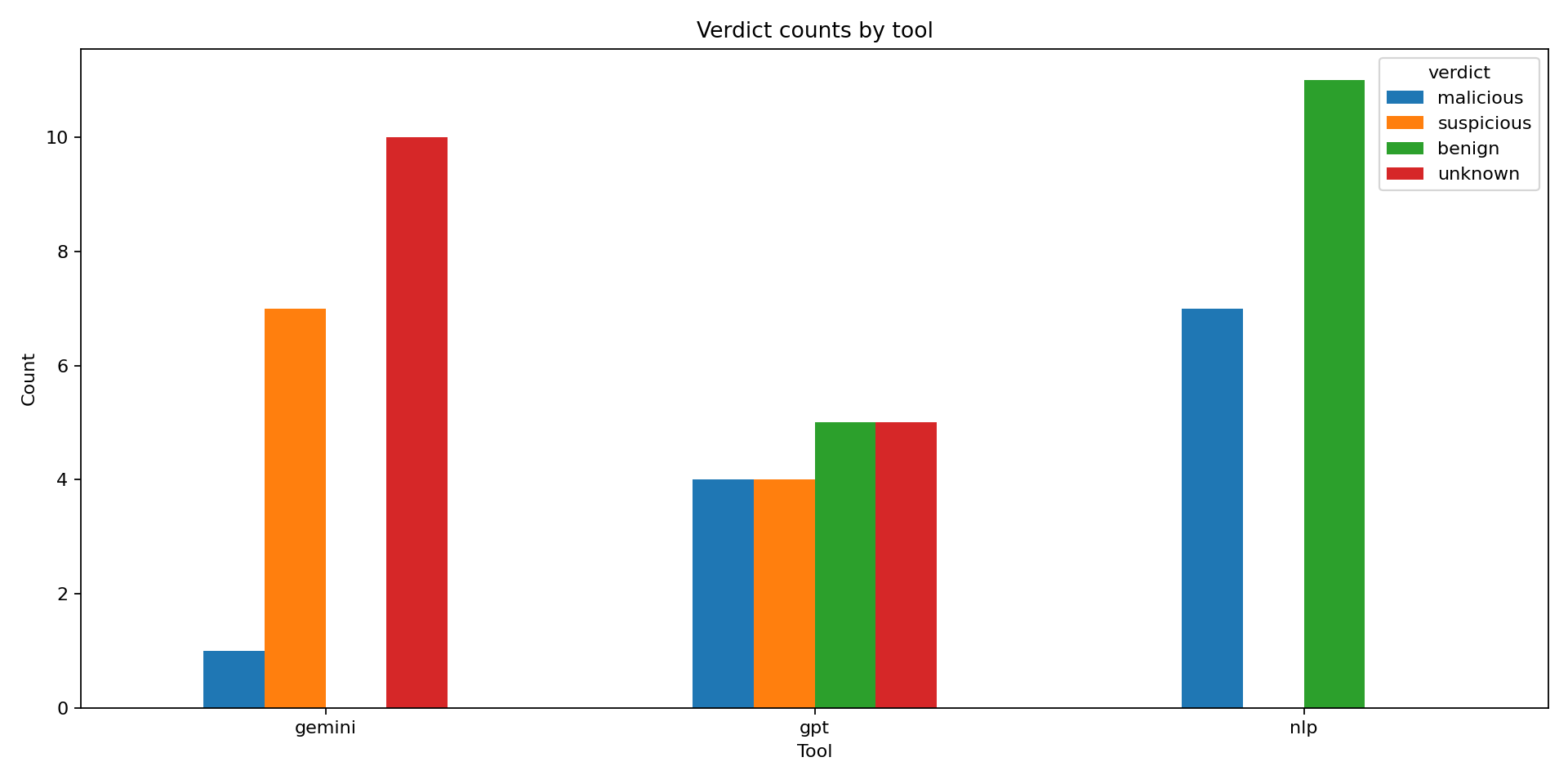}
    \caption{Evaluation results on Windows}
    \label{fig:resultsWin}
\end{figure}

\textbf{Comparison with LLMs and VirusTotal}.
The comparative analysis highlights that LLM-based memory forensics and VirusTotal address fundamentally different forensic problems.
VirusTotal primarily operates on static artifacts and reputation signals, relying on antivirus engine consensus, file hashes, and known indicators.
As a result, VirusTotal provides limited insight into runtime behavior and does not expose memory-resident artifacts such as injected code, live threads, kernel callbacks, or transient filesystem activity. Conversely, LLM-based analysis of Windows memory dumps is capable of extracting a broader and more diverse set of Indicators of Compromise directly from RAM.
These include evidence related to process injection, anomalous memory regions, filesystem interactions (temporary files, caches, databases), registry modifications, and kernel-level structures.
Our experiments show that LLMs consistently extract more IoCs than VirusTotal, particularly those related to dynamic behavior and system-level manipulation. Between the two LLMs, ChatGPT demonstrates more stable behavior in terms of verdict consistency and explanation quality, while Gemini shows higher variance and a greater tendency toward conservative or unknown classifications.
This difference is reflected in the lower agreement scores between Gemini and the other approaches. The quantitative results confirm that the three approaches produce substantially different verdict distributions.
The heuristic NLP classifier tends to polarize results toward benign or malicious based on explicit patterns, whereas LLMs exhibit a more nuanced distribution across malicious, suspicious, benign, and unknown categories. Pairwise agreement analysis further shows limited overlap between VirusTotal and LLM-based verdicts, indicating that reputation-based detection and memory forensics capture complementary aspects of malicious behavior. Higher agreement is observed between ChatGPT and the NLP approach, suggesting that rule-based IoC extraction partially aligns with LLMs reasoning when analyzing memory artifacts.

\begin{figure}
    \centering
    \includegraphics[width=0.6\linewidth]{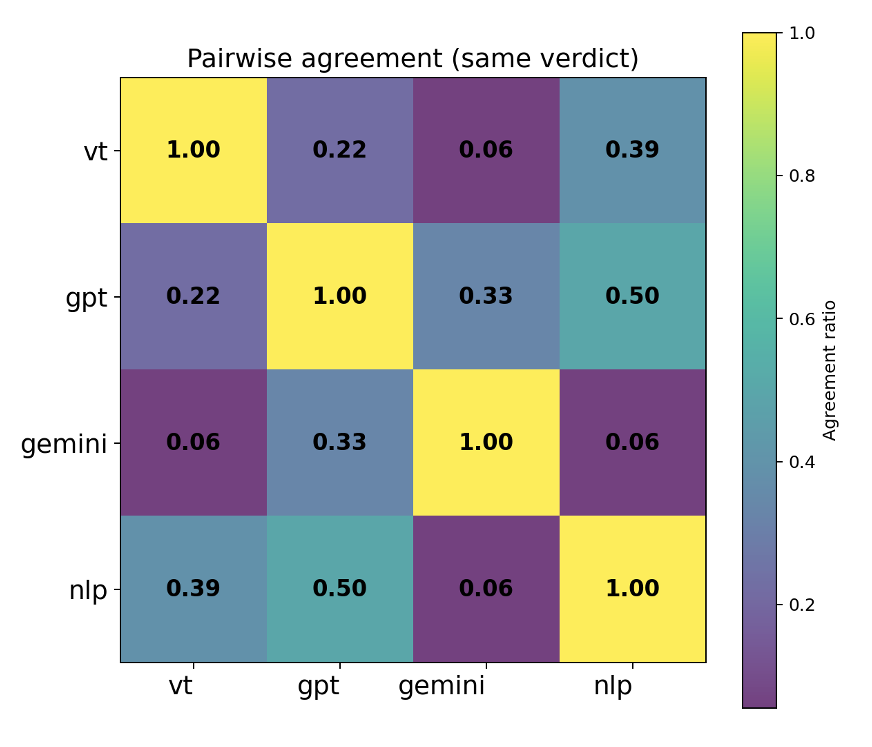}
    \caption{Comparison results on Windows}
    \label{fig:cmpWindows}
\end{figure}

Importantly, disagreement does not indicate incorrectness but rather reflects the fact that each method targets a distinct forensic layer: 
VirusTotal focuses on known malware signatures, NLP emphasizes explicit textual indicators, and LLMs synthesize heterogeneous memory evidence into higher-level forensic interpretations.

\label{sec:resultswin}

\subsection{Android}
Since the target device runs an Android system on a Linux kernel, the correct Linux plugin and matching profile must be selected to properly interpret kernel structures. However, available Linux plugins (particularly across Volatility~2 and~3) are limited, and kernel-dependent challenges make full RAM forensics less explored in prior work. Recent works such as {\tt VolMemDroid} focus on similar Android-specific guidelines and rely on a Goldfish~4.4 kernel, consistent with Android~9 (API~28). {\tt VolMemDroid} extracts IoCs using {\tt awk} and regex and converts them into feature vectors for analysis. We adopted the same execution environment, also because Goldfish kernel is still used and adopted by different devices even if release many years ago. In contrast, our methodology parses {\tt Volatility}’s plugin outputs and leverages general-purpose LLMs (ChatGPT-4o-mini and Gemini~2.0-flash-lite) to interpret results and generate human-readable summaries. This approach allows the automatic extraction of significant patterns (\ie IoCs, strings, and plugin-specific indicators) providing analysts with clear, interpretable reports.

\textbf{Memory Extraction}. The complete RAM extraction requires more resources because of the kernel recompilation. At the same time, the full RAM analysis gives more details on the effects of the target \APK in the whole system, for example monitoring written/accessed files and network connection. For this reason, we extracted the outputs of the selected {\tt Volatility} plugins (Table~\ref{tab:volplugins}) for each dump collected over time.

For each \APK we dumped $5$ times during the first $60$ seconds of execution (\ie every $10$ seconds which is also the time our machine needs to dump the memory of $2$GB, notably not corresponding to current real physical Android devices). We considered only a restricted subset of \APK, \ie dropper and fake \APK (\eg {\tt Netflix}, {\tt Twitter}, {\tt Whatsapp}, {\tt Instagram}, {\tt TikTok}, {\tt Facebook}) compared with the real official version. It is fundamental to recognize fake applications that could contain malicious behaviors while mimiking real legitimate applications not recognizable by users. One example is the recent Android stegomalware \cite{Soi_ITASEC25, Soi_IHMMSEC25, dellorco_pmcj25}.

\begin{figure}
    \centering
    \includegraphics[width=0.6\linewidth]{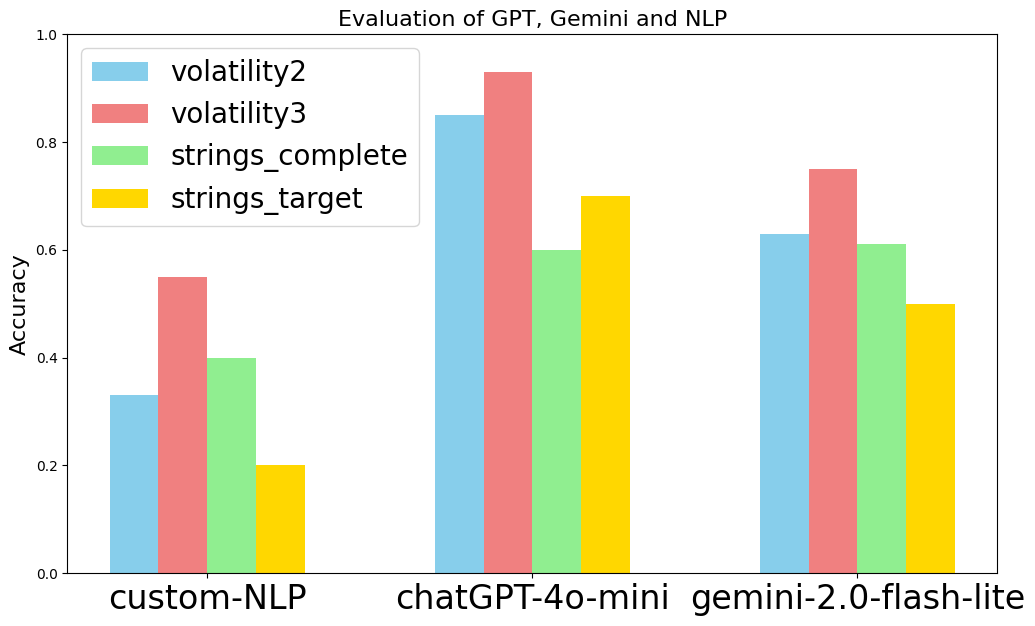}
    \caption{Comparison of the Android evaluation accuracy of the three models (from left to right: custom NLP, chatGPT-4o-mini, gemini-2.0-flash-lite) with the four different analysis (from left to right: volatility2, volatility3, strings extracted from the complete dump with LiME, and with the target dump with Frida}
    \label{fig:comparison}
\end{figure}

\textbf{Memory Analysis, Interpretability and Explanation (Android focus)}. As shown in Figure~\ref{fig:comparison}, {\tt chatGPT} is strictly performant in analysing the output of {\tt Volatility} and giving a human-readable description but also classifying them correctly, with an average confidence (\ie accuracy) of $0.93$ for Volatility~3. The LLM can classify the dumps in goodware/malware by also providing an explanation on the string pattern and its filename that lead to such classification (\eg malware because of the process \textit{com.quasar.bistocook2} found in linux.pstree), as shown in Table~\ref{tab:gpt_gemini} with a restricted dataset of fake \APK. Similarly, the \NLP algorithm extracts specific IoC based on regex and classifies the \APK according to the content of the memory dump. We made our own \NLP algorithm to parse both {\tt Volatility} outputs and the extracted strings both from the full RAM methodology (\ie {\tt LiMe}) and the target-process (\ie {\tt Fridump}), as displayed in Figure~\ref{fig:comparison}. The NLP component uses rule-based regular expressions and entropy checks to detect domain-specific forensic clues, producing interpretable feature vectors to classify memory-related text as malicious, suspicious, benign, or unknown. It is a knowledge-driven, explainable system that avoids statistical or embedding-based models.
 
Moreover, the models recognize the differences between fake and real \APK by highlighting differences when asked (Table~\ref{tab:gpt_gemini}). In particular, fake \APK can be detected because of the presence of IoCs regarding repacking, which is the technique to falsify \APK starting from existing ones \cite{Ruggia24_ASIACCS}. Moreover, thanks to the plugin {\tt malfind}, it can retrieve injected malicious payloads, especially when comparing fake samples against legitimate counterparts. Both LLMs, {\tt chatGPT} and {\tt Gemini}, can identify the name of the package exhibiting malicious behavior.

Overall, \LLM are more efficient than our \NLP algorithm, plausibly due to the broader knowledge base of general-purpose models, as reflected in Figure~\ref{fig:comparison}. {\tt Volatility} analysis with respect to {\tt Fridump} is more complete and the LLM can better explain the effects in the whole system. On the other hand, {\tt Fridump} analysis allows a more focused understanding of the target application behavior, especially if analyzed over time, by capturing actions such as runtime permissions, API calls, encryption, and dynamically loaded code.

\begin{table}
\centering
\setlength{\tabcolsep}{3pt} 
\renewcommand{\arraystretch}{1.1} 
\begin{tabular}{p{1.7cm} p{3cm} p{3cm}}
\hline
\textbf{Fake APK} & \textbf{GPT Detection} & \textbf{Gemini Detection} \\
\hline
Netflix  & process name, malfind & process name, process entries \\
Twitter  & connected IP, process name & process name, malfind \\
Tiktok   & process name, hidden process, malfind & process name, hidden process, malfind \\
Facebook & process name, domain & manipulation, domain \\
\hline
\end{tabular}
\caption{Comparison of GPT and Gemini string-based detections across a subset of Android fake applications}
\label{tab:gpt_gemini}
\end{table}

\begin{figure}
    \centering
    \includegraphics[width=0.6\linewidth]{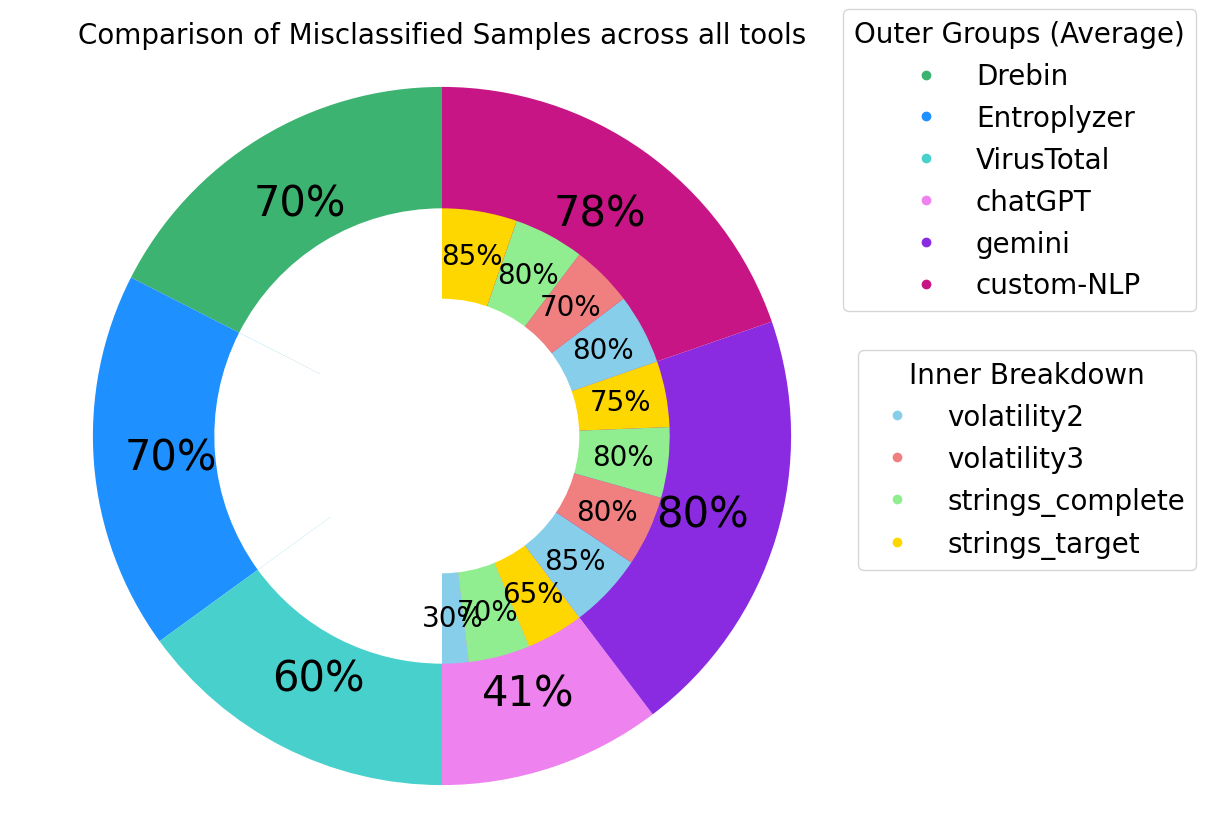}
    \caption{Average percentage of misclassified samples for the different Android detectors, \ie Drebin, Entroplyzer, VirusTotal (notably average percentage of AV engines that fail to detect on VT) and the LLM-RAM approach with the different LLMs and RAM analysis methodologies (volatility2, volatility3, strings from the complete dump or the target process dump)}
    \label{fig:camparison_sota}
\end{figure}

\textbf{Comparative Analysis}. Finally, we compared our memory forensics analysis with current literature tools, \ie {\tt VirusTotal}, {\tt Drebin} and {\tt Entroplyzer}. First of all, {\tt Drebin} and {\tt Entroplyzer} could analyze only some of them due to, respectively, anti-static analysis measures for anti-repacking in Androguard and anti-dynamic analysis for function hooking protection with Frida, as shown in Figure~\ref{fig:camparison_sota}.
The remaining \APK, less than $30$\% were successfully detected by the tools. With respect to {\tt VirusTotal} we compared the extracted IoC. Our memory forensics approach could find all IoC retrieved by VirusTotal, except for some IP addresses and domains that have been disabled. Additionally, the memory forensics approach could detect more IoCs for example regarding threads, file management (\eg db, tmp, cache files), filesystem interaction, syscalls, native libraries, hidden modules, as highlighted in Table~\ref{tab:ioc-comparison}. 

\begin{table}
\centering
\footnotesize
\setlength{\tabcolsep}{6pt}
\renewcommand{\arraystretch}{1.15}
\begin{tabularx}{\linewidth}{@{}>{\raggedright\arraybackslash}X>{\centering\arraybackslash}p{2.5cm}>{\centering\arraybackslash}p{3.5cm}@{}}
\toprule
\textbf{IoC / Evidence Category} & \textbf{VirusTotal} & \textbf{LLM\textendash RAM (Memory Forensics)} \\
\midrule
URLs/IPs/Domains                 & \cmark & \cmark \\
Bundled files / archives              & \cmark & \cmark \\
Files / artifacts on disk             & \xmark & \cmark \\
Executable code / processes           & \xmark & \cmark \\
System \& kernel structures           & \xmark & \cmark \\
User activity / environment           & \xmark & \cmark \\
Malware / manipulation indicators     & \xmark & \cmark \\
Device \& memory interfaces           & \xmark & \cmark \\
Tracing events      & \xmark & \cmark \\
\bottomrule
\end{tabularx}
\caption{Comparison of IoC categories supported by VirusTotal vs.\ Android LLM\textendash RAM.}
\label{tab:ioc-comparison}
\end{table}

\label{sec:resultsandro}

\section{Conclusions}
In this work, we propose a cross-platform memory analysis and interpretation methodology applicable to both Windows and Android systems. We introduce an analysis framework for malware detection and explanation based on the integration of general-purpose LLMs with volatile memory forensics. Given either structured outputs produced by {\tt Volatility} from Windows and Android full RAM images, or unstructured textual artifacts (\eg extracted strings) from memory dumps, the proposed approach enables human-readable interpretation, automatic extraction of Indicators of Compromise (IoCs), and explainable malware classification.

The text-based analysis (\ie string extraction and semantic interpretation) is applicable to both complete system memory snapshots and target-process memory regions, and is therefore suitable for desktop-class OS as well as mobile environments. Through a comparative evaluation of these acquisition strategies, we show that full RAM analysis provides a more comprehensive view of system-wide attack effects on the OS (particularly in Windows hosts), while target-process memory analysis can expose a higher density of application-specific IoCs, particularly in mobile scenarios.

We generate analyst-oriented reports in which the LLM explains why the examined memory artifacts derived from {\tt Volatility} plugin outputs or string-based analysis, should be considered benign, suspicious, or malicious, and explicitly links each classification decision to the extracted IoCs. The results are compared against established tools such as VirusTotal, demonstrating that memory-centric analysis reveals substantially richer forensic evidence than file-based reputation services alone, while remaining complementary to them.

We also show that a human-in-the-loop approach, supported by general-purpose LLMs during Android kernel-assisted memory acquisition and analysis, can significantly reduce setup complexity while preserving transparency and reproducibility, as demonstrated by the release of the compiled kernel artifacts and build metadata used in our experiments.

This study examines the applicability of Artificial Intelligence as a supportive tool in Digital Forensics analysis. \textbf{RQ1} investigates whether AI can assist critical phases of memory forensics, encompassing Android kernel recompilation for memory extraction and the analysis of acquired memory dumps through Volatility output interpretation, Indicators of Compromise (IoCs) extraction, malware classification, and result explanation \cite{sanna2025improvingcybercrimedetectiondigital}. \textbf{RQ2} focuses on the distinction between complete memory analysis and target-process analysis, where the former emphasizes operating system–level interactions and system-wide behavior, while the latter enables a more granular inspection of application-specific activities. \textbf{RQ3} evaluates the contribution of general-purpose LLMs to malware detection, assessing their capability to perform knowledge-driven IoC extraction and to generate human-readable explanations that enhance the interpretability of classification outcomes for forensic analysts.

While the proposed approach demonstrates promising results, several aspects warrant further investigation. The use of general-purpose LLMs is exploratory and primarily aimed at enhancing the interpretability of memory forensics outputs; a more extensive analysis of response consistency and reliability would be necessary to fully assess their suitability for forensic applications. Additionally, LLMs to be used in real Digital Forensics investigations must protect user data and commercial LLMs do not guarantee it, especially as data in memory dumps is in plain text. Hence, one solution would be the use of offline LLMs but we need specific data for training and ad-hoc tests on their performances must be conducted in the future. 

The framework builds upon well-established tools and techniques, and its contribution lies mainly in their integration within a unified workflow rather than in the development of novel acquisition mechanisms. The experimental evaluation is conducted on a limited set of samples and within an emulator-based environment using a single kernel configuration, which may restrict the generalization of the findings to other kernels, modern Generic Kernel Images, or physical devices. In addition, the requirement for kernel recompilation introduces practical constraints that should be considered when applying the approach in real-world forensic scenarios. Future work will focus on expanding the dataset, validating the methodology across diverse platforms, and further strengthening the evaluation of the LLM-based analysis component. The data used for this work is published online \footnote{\url{https://github.com/slsanna/LLM-chatbot-MFMA}}.
\label{sec:conclusions}

\section{Acknowledgments}
\label{sec:acks}
This work was partially supported by Project SERICS (PE00000014) under the NRRP MUR program funded by the EU - NGEU.

This work was carried out while Silvia Lucia Sanna was enrolled in the Italian National Doctorate on Artificial Intelligence run by Sapienza University of Rome in collaboration with the University of Cagliari.

\bibliographystyle{plain}
\bibliography{easychair}

@article{Breitinger2024DFRWS10yr,
  title={DFRWS EU 10-year review and future directions in Digital Forensic Research},
  author={Breitinger, Frank and Hilgert, Jan-Niclas and Hargreaves, Christopher and Sheppard, John and Overdorf, Rebekah and Scanlon, Mark},
  journal={Forensic Science International: Digital Investigation},
  volume={48},
  pages={301685},
  year={2024},
  doi={10.1016/j.fsidi.2023.301685}
}

@misc{zhao2025surveylargelanguagemodels,
      title={A Survey of Large Language Models}, 
      author={Wayne Xin Zhao and Kun Zhou and Junyi Li and Tianyi Tang and Xiaolei Wang and Yupeng Hou and Yingqian Min and Beichen Zhang and Junjie Zhang and Zican Dong and Yifan Du and Chen Yang and Yushuo Chen and Zhipeng Chen and Jinhao Jiang and Ruiyang Ren and Yifan Li and Xinyu Tang and Zikang Liu and Peiyu Liu and Jian-Yun Nie and Ji-Rong Wen},
      year={2025},
      eprint={2303.18223},
      archivePrefix={arXiv},
      primaryClass={cs.CL},
      url={https://arxiv.org/abs/2303.18223}, 
}

@inproceedings{Mengqi24_ACM,
author = {Liu, Mengqi and M'Hiri, Faten},
title = {Beyond Traditional Teaching: Large Language Models as Simulated Teaching Assistants in Computer Science},
year = {2024},
isbn = {9798400704239},
publisher = {Association for Computing Machinery},
address = {New York, NY, USA},
url = {https://doi.org/10.1145/3626252.3630789},
doi = {10.1145/3626252.3630789},
booktitle = {Proceedings of the 55th ACM Technical Symposium on Computer Science Education V. 1},
pages = {743–749},
numpages = {7},
location = {Portland, OR, USA},
series = {SIGCSE 2024}
}

@misc{sanna2025improvingcybercrimedetectiondigital,
      title={Improving Cybercrime Detection and Digital Forensics Investigations with Artificial Intelligence}, 
      author={Silvia Lucia Sanna and Leonardo Regano and Davide Maiorca and Giorgio Giacinto},
      year={2025},
      eprint={2510.14638},
      archivePrefix={arXiv},
      primaryClass={cs.CR},
      url={https://arxiv.org/abs/2510.14638}, 
}

@article{gobel2023data,
  title={Data for digital forensics: Why a discussion on “how realistic is synthetic data” is dispensable},
  author={G{\"o}bel, Thomas and Baier, Harald and Breitinger, Frank},
  journal={Digital Threats: Research and Practice},
  volume={4},
  number={3},
  pages={1--18},
  year={2023},
  publisher={ACM New York, NY}
}

@article{Mombelli2024FAIRness,
  title={FAIRness in digital forensics datasets’ metadata – and how to improve it},
  author={Mombelli, Samuele and Lyle, James R. and Breitinger, Frank},
  journal={Forensic Science International: Digital Investigation},
  volume={48},
  pages={301681},
  year={2024},
  doi={10.1016/j.fsidi.2023.301681}
}

@inproceedings{Franzen22_ACM,
author = {Franzen, Fabian and Holl, Tobias and Andreas, Manuel and Kirsch, Julian and Grossklags, Jens},
title = {Katana: Robust, Automated, Binary-Only Forensic Analysis of Linux Memory Snapshots},
year = {2022},
isbn = {9781450397049},
publisher = {Association for Computing Machinery},
address = {New York, NY, USA},
url = {https://doi.org/10.1145/3545948.3545980},
doi = {10.1145/3545948.3545980},
booktitle = {Proceedings of the 25th International Symposium on Research in Attacks, Intrusions and Defenses},
pages = {214–231},
numpages = {18},
location = {Limassol, Cyprus},
series = {RAID '22}
}

@book{Carvey2007,
  author    = {Harlan Carvey},
  title     = {Windows Forensic Analysis},
  publisher = {Syngress},
  year      = {2007}
}

@inproceedings{Schuster2006,
  author    = {Andreas Schuster},
  title     = {Searching for Processes and Threads in Microsoft Windows Memory Dumps},
  booktitle = {Digital Investigation},
  year      = {2006}
}

@article{Okolica2011,
  author  = {James Okolica and Gilbert Peterson},
  title   = {Windows Operating System Agnostic Memory Analysis},
  journal = {Digital Investigation},
  year    = {2011}
}

@inproceedings{Gianni2016,
  author    = {R. Gianni and others},
  title     = {Memory Forensics for Advanced Malware Detection},
  booktitle = {ACSAC},
  year      = {2016}
}

@inproceedings{Sylve2012,
  author    = {John Sylve and Andrew Case and Lodovico Marziale and Golden G. Richard},
  title     = {Acquisition and Analysis of Volatile Memory from Android Devices},
  booktitle = {DFRWS},
  year      = {2012}
}

@article{Case2013,
  author  = {Andrew Case and Jamie Levy},
  title   = {Automated Memory Forensics},
  journal = {Digital Investigation},
  year    = {2013}
}

@article{Afonso2016,
  author  = {Afonso, V. and others},
  title   = {Identifying Android Malware Using Dynamically Obtained Features},
  journal = {NDSS},
  year    = {2016}
}

@article{Garcia2018,
  author  = {Sebastián García and others},
  title   = {An Empirical Study of Machine Learning for Cybersecurity},
  journal = {IEEE Security \& Privacy},
  year    = {2018}
}

@article{Apruzzese2022,
  author  = {Giovanni Apruzzese and others},
  title   = {Machine Learning for Cybersecurity: A Systematic Review},
  journal = {ACM Computing Surveys},
  year    = {2022}
}

@article{Song2018,
  author  = {Song, Y. and others},
  title   = {Entropy-Based Detection of Packed Malware},
  journal = {SIGSAC},
  year    = {2018}
}

@article{Ribeiro2016,
  author  = {Marco Ribeiro and Sameer Singh and Carlos Guestrin},
  title   = {Why Should I Trust You? Explaining the Predictions of Any Classifier},
  journal = {KDD},
  year    = {2016}
}

@article{Guidotti2019,
  author  = {Riccardo Guidotti and others},
  title   = {A Survey of Explainable Artificial Intelligence},
  journal = {ACM Computing Surveys},
  year    = {2019}
}

@inproceedings {Ali-Gombe19_RAID,
author = {Aisha Ali-Gombe and Sneha Sudhakaran and Andrew Case and Golden G. Richard III},
title = {{DroidScraper}: A Tool for Android {In-Memory} Object Recovery and Reconstruction},
booktitle = {22nd International Symposium on Research in Attacks, Intrusions and Defenses (RAID 2019)},
year = {2019},
isbn = {978-1-939133-07-6},
address = {Chaoyang District, Beijing},
pages = {547--559},
url = {https://www.usenix.org/conference/raid2019/presentation/ali-gombe},
publisher = {USENIX Association},
month = sep
}

@conference{Zhang18_CPS,
	author = {Zhang, Junfu and Chengyuan, E. and Hu, Aiqun},
	title = {A method of android application forensics based on heap memory analysis},
	year = {2018},
	journal = {ACM International Conference Proceeding Series},
	doi = {10.1145/3207677.3277934},
	url = {https://www.scopus.com/inward/record.uri?eid=2-s2.0-85056819842&doi=10.1145%2f3207677.3277934&partnerID=40&md5=c42d0dd2e9a55805b2105bc27b5990f1},
	type = {Conference paper},
	publication_stage = {Final},
	source = {Scopus},
	note = {Cited by: 0}
}

@INPROCEEDINGS{Keyes21_RDAAPS,
  author = {Keyes, David Sean and Li, Beiqi and Kaur, Gurdip and Lashkari, Arash Habibi and Gagnon, Francois and Massicotte, Frédéric},
  title = {EntropLyzer: Android Malware Classification and Characterization Using Entropy Analysis of Dynamic Characteristics},
  year = {2021},
  booktitle = {2021 Reconciling Data Analytics, Automation, Privacy, and Security: A Big Data Challenge (RDAAPS)},
  volume = {},
  number = {},
  pages = {1-12},
  doi = {10.1109/RDAAPS48126.2021.9452002}
}

@article{Soi_ITASEC25,
  author = {Soi, Diego and Sanna, Silvia Lucia and Liguori, Angelica and Zuppelli, Marco and Regano, Leonardo and Maiorca, Davide and Caviglione, Luca and Manco, Giuseppe and Giacinto, Giorgio},
  title = {{On the Feasibility of Android Stegomalware: A Detection Study}},
  year = {2025},
  journal = {{Italian Conference on CyberSecurity}},
  note = {{https://ceur-ws.org/Vol-3962/paper9.pdf}}
}

@article{Demontis19_IEEE,
  author = {Demontis, Ambra and Melis, Marco and Biggio, Battista and Maiorca, Davide and Arp, Daniel and Rieck, Konrad and Corona, Igino and Giacinto, Giorgio and Roli, Fabio},
  title = {Yes, Machine Learning Can Be More Secure! A Case Study on Android Malware Detection},
  year = {2019},
  journal = {IEEE Trans. Dependable Secur. Comput.},
  publisher = {IEEE Computer Society Press},
  pages = {711–724},
  doi = {10.1109/TDSC.2017.2700270},
  month = {jul},
  numpages = {14}
}

@ARTICLE{Bellizzi_Access22,
  author={Bellizzi, Jennifer and Vella, Mark and Colombo, Christian and Hernandez-Castro, Julio},
  journal={IEEE Access}, 
  title={Responding to Targeted Stealthy Attacks on Android Using Timely-Captured Memory Dumps}, 
  year={2022},
  volume={10},
  number={},
  pages={35172-35218},
  doi={10.1109/ACCESS.2022.3160531}}

@InProceedings{Bellizzi_SIP21,
  author = {Bellizzi, Jennifer
and Vella, Mark
and Colombo, Christian
and Hernandez-Castro, Julio},
  title = {Real-Time Triggering of Android Memory Dumps for Stealthy Attack Investigation},
  year = {2021},
  booktitle = {Secure IT Systems},
  publisher = {Springer International Publishing},
  pages = {20--36},
  isbn = {978-3-030-70852-8},
  editor = {Asplund, Mikael
and Nadjm-Tehrani, Simin},
  address = {Cham}
}

@Article{Bellizzi_JCP23_vedrando,
  author = {Bellizzi, Jennifer and Losiouk, Eleonora and Conti, Mauro and Colombo, Christian and Vella, Mark},
  title = {VEDRANDO: A Novel Way to Reveal Stealthy Attack Steps on Android through Memory Forensics},
  year = {2023},
  journal = {Journal of Cybersecurity and Privacy},
  volume = {3},
  number = {3},
  pages = {364--395},
  doi = {10.3390/jcp3030019},
  url = {https://www.mdpi.com/2624-800X/3/3/19},
  issn = {2624-800X}
}

@misc{virustotal,
  title = {VirusTotal},
  howpublished = {\url{https://www.virustotal.com}}
}

@inproceedings{drebin,
  author = {Arp, Daniel and Spreitzenbarth, Michael and Hubner, Malte and Gascon, Hugo and Rieck, Konrad},
  title = {DREBIN: Effective and Explainable Detection of Android Malware in Your Pocket.},
  year = {2014},
  booktitle = {NDSS},
  publisher = {The Internet Society},
  url = {http://dblp.uni-trier.de/db/conf/ndss/ndss2014.html#ArpSHGR14},
  ee = {https://www.ndss-symposium.org/ndss2014/drebin-effective-and-explainable-detection-android-malware-your-pocket}
}

@article{bellizzi2022jitmf,
  author = {Bellizzi, Jennifer and Bruno, Alessandro and Martinelli, Fabio},
  title = {JIT-MF: Just-in-Time Memory Forensics for Android Devices},
  year = {2022},
  journal = {Digital Investigation},
  volume = {42},
  pages = {301450},
  url = {https://arxiv.org/abs/2103.04891}
}

@article{Bhusal_ACM25,
  author = {Bhusal, Dipkamal and Rastogi, Nidhi},
  title = {Adversarial Patterns: Building Robust Android Malware Classifiers},
  year = {2025},
  journal = {ACM Comput. Surv.},
  publisher = {Association for Computing Machinery},
  volume = {57},
  number = {8},
  doi = {10.1145/3717607},
  url = {https://doi.org/10.1145/3717607},
  issn = {0360-0300},
  address = {New York, NY, USA},
  month = {mar},
  articleno = {192},
  issue_date = {August 2025},
  numpages = {34}
}

@inproceedings{Soi_IHMMSEC25,
  author = {Soi, Diego and Sanna, Silvia Lucia and Benedetti, Giacomo and Liguori, Angelica and Regano, Leonardo and Caviglione, Luca and Giacinto, Giorgio},
  title = {Analysis and Detection of Android Stegomalware: the Impact of the Loading Stage},
  year = {2025},
  booktitle = {Proceedings of the 2025 ACM Workshop on Information Hiding and Multimedia Security},
  publisher = {Association for Computing Machinery},
  pages = {35–45},
  doi = {10.1145/3733102.3733122},
  url = {https://doi.org/10.1145/3733102.3733122},
  isbn = {9798400718878},
  series = {IH\&MMSEC '25},
  address = {New York, NY, USA},
  location = {},
  numpages = {11}
}

@article{dellorco_pmcj25,
  author = {Danilo Dell’Orco and Giorgio Bernardinetti and Giuseppe Bianchi and Alessio Merlo and Alessandro Pellegrini},
  title = {Would you mind hiding my malware? Building malicious Android apps with StegoPack},
  year = {2025},
  journal = {Pervasive and Mobile Computing},
  volume = {111},
  pages = {102060},
  doi = {https://doi.org/10.1016/j.pmcj.2025.102060},
  url = {https://www.sciencedirect.com/science/article/pii/S1574119225000495},
  issn = {1574-1192}
}

@article{Aligombe_DFRWS23_crgbmem,
  author = {Aisha Ali-Gombe and Sneha Sudhakaran and Ramyapandian Vijayakanthan and Golden G. Richard},
  title = {cRGB\_Mem: At the intersection of memory forensics and machine learning},
  year = {2023},
  journal = {Forensic Science International: Digital Investigation},
  volume = {45},
  pages = {301564},
  doi = {https://doi.org/10.1016/j.fsidi.2023.301564},
  url = {https://www.sciencedirect.com/science/article/pii/S2666281723000732},
  issn = {2666-2817}
}

@inproceedings{Nataraj_VizSec11,
  author = {Nataraj, L. and Karthikeyan, S. and Jacob, G. and Manjunath, B. S.},
  title = {Malware images: visualization and automatic classification},
  year = {2011},
  booktitle = {Proceedings of the 8th International Symposium on Visualization for Cyber Security},
  publisher = {Association for Computing Machinery},
  doi = {10.1145/2016904.2016908},
  url = {https://doi.org/10.1145/2016904.2016908},
  isbn = {9781450306799},
  series = {VizSec '11},
  address = {New York, NY, USA},
  articleno = {4},
  location = {Pittsburgh, Pennsylvania, USA},
  numpages = {7}
}

@ARTICLE{Malhotra2024_Virology,
  author = {Malhotra, Vrinda and Potika, Katerina and Stamp, Mark},
  title = {A comparison of graph neural networks for malware classification},
  year = {2024},
  journal = {{Journal of Computer Virology and Hacking Techniques}},
  volume = {20},
  number = {1},
  pages = {53 – 69},
  doi = {10.1007/s11416-023-00493-y}
}

@article{Khalid_volmemdroid_ESA24,
  author = {Saneeha Khalid and Faisal Bashir Hussain},
  title = {VolMemDroid—Investigating android malware insights with volatile memory artifacts},
  year = {2024},
  journal = {Expert Systems with Applications},
  volume = {253},
  pages = {124347},
  doi = {https://doi.org/10.1016/j.eswa.2024.124347},
  url = {https://www.sciencedirect.com/science/article/pii/S0957417424012132},
  issn = {0957-4174}
}

@misc{504ensicsLabsLiME,
  author = {504ensicsLabs},
  title = {LiME: Linux Memory Extractor (Loadable Kernel Module for volatile memory acquisition)},
  year = {2012},
  note = {GitHub repository},
  howpublished = {\url{https://github.com/504ensicsLabs/LiME}}
}

@misc{VolatilityFramework,
  author = {Volatility Foundation},
  title = {Volatility: The memory forensics framework},
  note = {Open source memory forensics tool},
  howpublished = {\url{https://www.volatilityfoundation.org/}}
}

@article{gaber_malware_2024,
  author = {Gaber, M.G. and Ahmed, M. and Janicke, H.},
  title = {Malware {Detection} with {Artificial} {Intelligence}: {A} {Systematic} {Literature} {Review}},
  year = {2024},
  journal = {ACM Computing Surveys},
  volume = {56},
  number = {6},
  doi = {10.1145/3638552},
  shorttitle = {Malware {Detection} with {Artificial} {Intelligence}}
}

@misc{Volatility,
  author = {{Volatility Foundation}},
  title = {{Volatility Framework}},
  url = {https://github.com/volatilityfoundation/volatility}
}

@misc{Frida,
  author = {{Frida documentation}},
  title = {{Frida}},
  year = {{}},
  url = {{https://frida.re/docs/home/}},
  note = {{Online; Accessed 02/2025}}
}

@article{mercaldo2021,
  author    = {Mercaldo, Francesco and Santone, Antonella},
  title     = {Audio signal processing for Android malware detection and family identification},
  journal   = {Journal of Computer Virology and Hacking Techniques},
  year      = {2021},
  volume    = {17},
  number    = {2},
  pages     = {139--152},
  doi       = {10.1007/s11416-020-00376-6},
  url       = {https://doi.org/10.1007/s11416-020-00376-6}
}

@article{kural2023,
  author={Kural, Oguz Emre and Kiliç, Erdal and Aksaç, Ceyda},
  journal={IEEE Access}, 
  title={Apk2Audio4AndMal: Audio Based Malware Family Detection Framework}, 
  year={2023},
  volume={11},
  number={},
  pages={27527-27535},
  doi={10.1109/ACCESS.2023.3258377}}

@misc{Daoudi21_Springer,
  author = {Jordan Daoudi and Tegawend\'{e} F. Bissyand{\'e} and Jacques Klein},
  title = {DexRay: Simple, Efficient and Effective Android Malware Detection via Bytecode Gray-Scale Imaging},
  year = {2021},
  note = {arXiv preprint},
  archiveprefix = {arXiv},
  eprint = {2108.}
}

@inproceedings{Ruggia24_ASIACCS,
author = {Ruggia, Antonio and Nisi, Dario and Dambra, Savino and Merlo, Alessio and Balzarotti, Davide and Aonzo, Simone},
title = {Unmasking the Veiled: A Comprehensive Analysis of Android Evasive Malware},
year = {2024},
isbn = {9798400704826},
publisher = {Association for Computing Machinery},
address = {New York, NY, USA},
url = {https://doi.org/10.1145/3634737.3637658},
doi = {10.1145/3634737.3637658},
booktitle = {Proceedings of the 19th ACM Asia Conference on Computer and Communications Security},
pages = {383–398},
numpages = {16},
location = {Singapore, Singapore},
series = {ASIA CCS '24}
}

@inproceedings{Ruggia24_AndroidEvasive,
author = {Ruggia, Antonio and Nisi, Dario and Dambra, Savino and Merlo, Alessio and Balzarotti, Davide and Aonzo, Simone},
title = {Unmasking the Veiled: A Comprehensive Analysis of Android Evasive Malware},
year = {2024},
isbn = {9798400704826},
publisher = {Association for Computing Machinery},
address = {New York, NY, USA},
url = {https://doi.org/10.1145/3634737.3637658},
doi = {10.1145/3634737.3637658},
booktitle = {Proceedings of the 19th ACM Asia Conference on Computer and Communications Security},
pages = {383–398},
numpages = {16},
location = {Singapore, Singapore},
series = {ASIA CCS '24}
}

@inproceedings{yang2019fridump,
  author = {Yang, Lin and Zhang, Bo and Wang, Yi},
  title = {Process-based Memory Extraction via Frida for Android Forensics},
  year = {2019},
  booktitle = {Proceedings of the DFRWS EU}
}

@manual{rekall2017,
  author = {{Rekall Project}},
  title = {Rekall Memory Forensic Framework},
  year = {2017},
  note = {\url{https://www.rekall-forensic.com}}
}


\appendix
\captionsetup{justification=centering}

\FloatBarrier
\begin{table*}
\centering
\footnotesize
\renewcommand{\arraystretch}{1.1}
\begin{tabular}{p{3.5cm}p{10cm}}
    \toprule
    \textbf{Windows Category} & \textbf{Volatility 3 Plugins} \\
    \midrule

    \textbf{Process \& Execution Analysis} &
    windows.pslist.PsList, windows.psscan.PsScan, windows.pstree.PsTree, windows.cmdline.CmdLine, windows.envars.Envars, windows.handles.Handles, windows.sessions.Sessions, windows.joblinks.JobLinks \\
    \midrule

    \textbf{Memory Injection \& Malware Techniques} &
    windows.malfind.Malfind, windows.hollowprocesses.HollowProcesses, windows.processghosting.ProcessGhosting, windows.malware.pebmasquerade.PebMasquerade, windows.malware.hollowprocesses.HollowProcesses, windows.malware.processghosting.ProcessGhosting \\
    \midrule

    \textbf{Loaded Modules \& Drivers} &
    windows.dlllist.DllList, windows.modules.Modules, windows.ldrmodules.LdrModules, windows.unloadedmodules.UnloadedModules, windows.driverscan.DriverScan, windows.drivermodule.DriverModule, windows.malware.drivermodule.DriverModule \\
    \midrule

    \textbf{Kernel \& Low-Level Structures} &
    windows.ssdt.SSDT, windows.kpcrs.KPCRs, windows.callbacks.Callbacks, windows.driverirp.DriverIrp, windows.timers.Timers \\
    \midrule

    \textbf{Memory Layout \& VAD Analysis} &
    windows.memmap.Memmap, windows.vadinfo.VadInfo, windows.vadwalk.VadWalk, windows.vadregexscan.VadRegExScan, windows.virtmap.VirtMap, windows.bigpools.BigPools \\
    \midrule

    \textbf{Persistence \& Autostart Mechanisms} &
    windows.registry.scheduled\_tasks.ScheduledTasks, windows.scheduled\_tasks.ScheduledTasks, windows.registry.userassist.UserAssist, windows.shimcachemem.ShimcacheMem, windows.amcache.Amcache, windows.registry.amcache.Amcache \\
    \midrule

    \textbf{Registry Analysis} &
    windows.registry.hivelist.HiveList, windows.registry.hivescan.HiveScan, windows.registry.printkey.PrintKey, windows.registry.getcellroutine.GetCellRoutine, windows.registry.certificates.Certificates \\
    \midrule

    \textbf{Filesystem \& Object Artifacts} &
    windows.filescan.FileScan, windows.mutantscan.MutantScan, windows.symlinkscan.SymlinkScan, windows.poolscanner.PoolScanner \\
    \midrule

    \textbf{Networking \& System Context} &
    windows.info.Info, windows.statistics.Statistics, windows.getsids.GetSIDs, windows.getservicesids.GetServiceSIDs, windows.svclist.SvcList, windows.svcscan.SvcScan, windows.svcdiff.SvcDiff \\
    \midrule

    \textbf{User Interface \& Session Objects} &
    windows.deskscan.DeskScan, windows.desktops.Desktops, windows.windowstations.WindowStations, windows.windows.Windows \\
    \midrule

    \textbf{Miscellaneous \& Special-Purpose} &
    windows.mbrscan.MBRScan, windows.truecrypt.Passphrase, windows.devicetree.DeviceTree, windows.strings.Strings \\
    \bottomrule
\end{tabular}
\caption[Windows Volatility 3 Plugin Categorization]{Categorization of Windows Volatility~3 plugins grouped by forensic analysis domain}
\label{tab:volplugins_windows}
\end{table*}

\FloatBarrier
\begin{table*}
\centering
\footnotesize
\renewcommand{\arraystretch}{1.1}
\begin{tabular}{p{3.3cm}p{5.5cm}p{5.5cm}}
    \toprule
    \textbf{Linux Category} & \textbf{Volatility 2 Plugins} & \textbf{Volatility 3 Plugins}\\
    \midrule
    \textbf{Process \& Execution Analysis} &
    linux\_pslist, linux\_psscan, linux\_pstree, linux\_paux, linux\_psenv, linux\_psxview, linux\_threads &
    linux.pslist.PsList, linux.psscan.PsScan, linux.pstree.PsTree, linux.psaux.PsAux, linux.pscallstack.PsCallStack, linux.ptrace.Ptrace, linux.kthreads.Kthreads \\
    \midrule
    \textbf{Kernel \& System Structures} &
    linux\_pidhashtable, linux\_iomem, linux\_lsmod, linux\_ldrmodules, linux\_library\_list, linux\_banner &
    linux.kallsyms.Kallsyms, linux.pidhashtable.PIDHashTable, linux.iomem.IOMem, linux.module\_extract.ModuleExtract, linux.lsmod.Lsmod, linux.modxview.Modxview, linux.library\_list.LibraryList \\
    \midrule
    \textbf{Malware \& Rootkit Detection} &
    linux\_malfind, linux\_hidden\_modules, linux\_check\_creds, linux\_check\_idt, linux\_check\_syscall, linux\_check\_modules, linux\_check\_tty, linux\_plthook &
    linux.malfind.Malfind, linux.hidden\_modules.Hidden\_modules, linux.check\_creds.Check\_creds, linux.check\_idt.Check\_idt, linux.check\_syscall.Check\_syscall, linux.check\_modules.Check\_modules, linux.check\_afinfo.Check\_afinfo, linux.netfilter.Netfilter, linux.keyboard\_notifiers.Keyboard\_notifiers, linux.ebpf.EBPF, linux.tty\_check.tty\_check \\
    \midrule
    \textbf{Filesystem \& Page Cache} &
    linux\_enumerate\_files, linux\_recover\_filesystem, linux\_mount, linux\_kernel\_opened\_files, linux\_lsof &
    linux.pagecache.Files, linux.pagecache.InodePages, linux.pagecache.RecoverFs \\
    \midrule
    \textbf{Networking \& Communication} &
    linux\_ifconfig, linux\_netstat, linux\_netscan, linux\_route\_cache, linux\_sk\_buff\_cache &
    linux.ip.Addr, linux.ip.Link, linux.sockstat.Sockstat, linux.netfilter.Netfilter \\
    \midrule
    \textbf{Tracing \& Instrumentation} &
    linux\_info\_regs &
    linux.tracing.ftrace.CheckFtrace, linux.tracing.perf\_events.PerfEvents, linux.tracing.tracepoints.CheckTracepoints \\
    \midrule
    \textbf{User Environment \& Capabilities} &
    linux\_bash, linux\_bash\_env, linux\_bash\_hash, linux\_dynamic\_env &
    linux.bash.Bash, linux.capabilities.Capabilities, linux.envars.Envars \\
    \midrule
    \textbf{Miscellaneous \& Device Interfaces} &
    linux\_truecrypt\_passphrase, linux\_dmesg &
    linux.kmsg.Kmsg, linux.graphics.fbdev.Fbdev, linux.vmaregexscan.VmaRegExScan, linux.vmayarascan.VmaYaraScan \\
    \bottomrule
\end{tabular}
\caption[Comparison of Volatility 2 and 3 Linux Plugins]{Comparison between Linux plugins in Volatility~2 and Volatility~3, grouped by functional typology}
\label{tab:volplugins}
\end{table*}   
\FloatBarrier

\FloatBarrier
\begin{table*}
\centering
\small
\setlength{\tabcolsep}{3pt}
\renewcommand{\arraystretch}{1.08}
\resizebox{\textwidth}{!}{%
\begin{tabularx}{\textwidth}{
    l
    p{0.16\textwidth}
    p{0.48\textwidth}
    >{\raggedright\arraybackslash}p{0.22\textwidth}
}
\toprule
\textbf{Level} & \textbf{Goal} & \textbf{Full prompt content (preamble + numbered tasks)} & \textbf{JSON keys (output)} \\
\midrule
\textbf{Plugin level} & Explain one plugin/file; extract IoCs &
\textit{System role:} “You are a forensic analyst.” Given the text file from a memory dump plugin, perform three tasks: 1) Briefly explain what the plugin/file output shows (describe plugin and summarize output; focus on findings). 2) Extract all observable, investigation-relevant Indicators of Compromise (IoCs) from the file. 3) Keep the explanation neutral and concise. &
\texttt{file\_summary}; \texttt{iocs[\{value,type,\-evidence\}]} \\[0.4em]
\textbf{Dump level} & Correlate files, infer verdict \& rationale &
\textit{System role:} “You are a forensic analyst.” You receive (1) multiple short per-file summaries and IoCs (JSON objects), and (2) truncated excerpts from the same dump (for context). \textit{Tasks:} 1) Produce one explanation for the dump (aggregate/interpret the content). 2) Extract all IoCs across everything provided. 3) Decide a \texttt{dump\_verdict} = \texttt{malicious|suspicious|benign|unknown} and explain why using IoCs only; provide \texttt{recommended\_next\_steps}. &
\texttt{summary}; \texttt{iocs}; \texttt{dump\_verdict}; \texttt{confidence}; \texttt{rationale}; \texttt{recommended\_next\-\_steps} \\[0.4em]
\textbf{App level} & Aggregate dump verdicts \& IoCs across volumes into a single app verdict &
\textit{No LLM prompt (deterministic aggregation).} \textit{Inputs:} per-dump results. \textit{Steps:} 1) Count dump verdicts \& list dumps by verdict; 2) aggregate IoCs \emph{(type,value,count,dumps,examples)} and keep top N (30–40); 3) decide \texttt{app\_verdict}: \texttt{malicious} if any malicious dumps, else \texttt{suspicious} if any suspicious, else \texttt{benign} if any benign, else \texttt{unknown}; 4) write recap + sections (app Verdict, Dump Verdicts, Top IoCs, Notes). &
\texttt{app\_verdict}; \texttt{verdict\_counts}; \texttt{by\_verdict[dumps]}; \texttt{top\_iocs[type,value,\-dumps]}; \texttt{summary/recap}; \texttt{notes} \\[0.35em]
\bottomrule
\end{tabularx}}
\caption{Full prompt hierarchy: from per-plugin and dump-level GPT analyses to App-level aggregation. \textit{Constraint:} Ignore all forensic/tooling artefacts (Volatility, LiME, plugin names, \texttt{insmod}/\texttt{modprobe}, tracebacks, warnings, errors); they must not appear as IoCs or influence verdicts.}
\label{tab:fullprompts}
\end{table*}
\FloatBarrier

\end{document}